\documentclass[
 reprint,
 amsmath,amssymb,
 aps,
]{revtex4-2}

\usepackage{graphicx}
\usepackage{bm}
\usepackage{lineno}
\usepackage{siunitx}
\usepackage{url}
\usepackage{hyperref}

\usepackage{dcolumn}
\newcolumntype{d}[1]{D{.}{.}{#1}}

\usepackage[ruled,vlined]{algorithm2e}
\usepackage{algorithmic}
\begin{document}

\title{Statistical mechanics for organic mixed conductors: phase transitions in a lattice gas}

\author{Lukas M. Bongartz}
\affiliation{
 Institute for Applied Physics, Technische Universit\"at Dresden, 01187 Dresden, Germany
}

\begin{abstract}
Organic mixed conductors (OMCs) represent a promising class of materials for applications in bioelectronics, physical computing, and thermoelectrics. Rather unparalleled, OMCs feature dynamics spanning multiple length and time scales, involving an intricate coupling between electronic, ionic, and mass transport. These characteristics set them notably apart from traditional semiconductors and hinder the description by conventional semiconductor theory. In this work, we approach the charge carrier modulation of OMCs using statistical mechanics. We discuss OMCs from a thermodynamic perspective and contrast them with established semiconductor materials, highlighting key differences in their collective charge carrier dynamics. This motivates our description of OMCs as a lattice gas, which we analyze within the grand canonical ensemble. The model exhibits a first-order phase transition analogous to a classical vapor--liquid transition, governed by temperature and chemical potential. In doing so, it captures the formation of distinct low- and high-density carrier phases, consistent with recently reported experimental observations. It also illustrates how metastability near the phase boundary can give rise to history-dependent characteristics in device operation, a similarly well-reported effect in OMC transistors. This work is intended as a simple motivation for studying OMCs through the lens of statistical mechanics, offering a more natural description than traditional semiconductor models developed for materials of fundamentally distinct character.
\end{abstract}

\maketitle

\section{Introduction}

Organic mixed conductors (OMCs) -- semiconducting polymers that allow for both ion and electron transport -- represent a novel class of electronic materials, distinct from conventional inorganic systems like silicon \cite{paulsen2020organic}. While traditional semiconductors are generally operated (switched between conductive and non-conductive states) using the field effect, OMCs are controlled electrochemically. In this mode, the dynamic introduction and retrieval of dopants into the semiconductor become a central part of device operation, as opposed to the singular introduction of dopants during fabrication in legacy systems.

While OMC technology is still in its infancy and fundamental research stage, it is beginning to crystallize as a promising candidate to complement application layers inaccessible to silicon et al. These include, in particular, bioelectronics such as neural interfaces and bio-inspired computing \cite{gkoupidenis2024organic}. The archetypal OMC device is the organic electrochemical transistor (OECT) \cite{Rivnay2018}. In this device architecture, the semiconductor channel is coupled to a gate electrode via an electrolyte. The electrolyte acts as an ion reservoir, donating dopants into the polymer semiconductor mesh, where they compensate the (n- or p-type) electronic charge carriers hosted within the OMC's $\pi$-system. As in conventional transistor devices, source and drain contacts probe the channel conductivity in terms of a drain current, with the key difference that carriers are driven throughout the entire volume of the channel, rather than being confined to the surface. We want to stress that OMCs represent one of the most diverse materials classes developed to date, where minor modifications in the molecular structure or material processing can manifest in profound differences on the microscale (e.g., morphology, carrier mobility, mass and charge capacity) \cite{wang2024designing}. We here take a holistic view on OMCs, intentionally abstracting away from microscopic details to focus on the general characteristics of charge carrier dynamics. We thus consider generalized model systems of both silicon-like and carbon-based polymer semiconductors. This viewpoint already allows us to identify several key distinctions between the physics governing each.

Most notably, OMC operation features extremely high charge carrier densities ($\sim 10^{20}$--$10^{21}$\,cm$^{-3}$), orders of magnitude beyond what is typical in inorganic semiconductors and approaching a 1:1 population of electronic charges on the molecular hosting sites \cite{frisbie2026charge, he2022sub, tjhe2024non}. Due to the disordered morphology, charge transport is often of a hopping nature, rather than band-like \cite{fratini2020charge, ihnatsenka2015understanding, kim2012thermoelectric}. As carbon-based systems, OMCs also feature low dielectric constants \cite{warren2023molecular, ortstein2021band}, which amplify both Coulomb and higher-order multipole interactions. While the electrolytic environment partially screens long-range Coulomb interactions (Debye screening), short-range correlations remain significant. A simple order-of-magnitude estimate using archetypal material parameters (Appendix~\ref{app:estimate_coupling}) suggests that carrier--carrier coupling in OMCs exceeds that in silicon-based systems by several $k_\mathrm{B}T$, across the entire regime of relevant carrier densities ($10^{16}$--$10^{19}\,\text{cm}^{-3}$ for silicon; $10^{18}$--$10^{21}\,\text{cm}^{-3}$ for OMCs). This has important consequences for charge carrier dynamics, causing deviations from standard frameworks such as Miller--Abrahams hopping transport \cite{vukmirovic2010carrier}.

Here, we focus on the charge carrier density and its modulation by an external bias, while leaving the incorporation of transport phenomena for a follow-up publication. Several works have shown the impact of particle interactions in OMC operation, such as electron--electron \cite{tjhe2024non, koopmans2021carrier}, electron--phonon \cite{spano2014h}, and electron--dopant modes \cite{tjhe2024non, koopmans2020electrical, bongartz2025electron}. Collectively, these give rise to a dynamic density of states that evolves with the carrier density itself and contribute to a heterogeneous doping process. They reveal an intricate configurational landscape involving the coupling between electronic charge carriers, mobile ionic dopants, molecular vibrations, and morphology. Two recent reports on an established OMC serve as an illustrative case study: Wu et al. provided direct experimental evidence that an electrochemically doped OMC forms $\sim$10\,nm domains of distinct electronic character \cite{wu2024bridging}, whose growth is consistent with Model-B-type coarsening \cite{hohenberg1977theory}. This non-equilibrium phase separation is tied to a strong coupling between carrier populations and structural rearrangements, linking it to Cahn--Hilliard--like dynamics at the mesoscale \cite{cahn1958free}. Using the same OMC with a strongly interacting electrolyte, Bongartz et al. demonstrated OECTs with persistent hysteresis on experimental timescales, thermodynamically ascribed to a phase separation into domains of high and low doping degree \cite{bongartz2024bistable}. 

These observations converge on a common underlying theme: a non-uniform stabilization of charge carriers, which suggests a connection to our coupling estimates in Appendix~\ref{app:estimate_coupling}. In inorganic semiconductors like silicon, charge carriers remain strongly delocalized, weakly interacting, and efficiently screened by fast electronic polarization, which suppresses correlation effects. This leads to the well-known description as a quasi-free electron gas governed by effective-mass physics. In polymer semiconductors, the situation is fundamentally different: while mobile ions in the electrolytic environment effectively screen long-range Coulomb repulsion, strong electron–phonon coupling leads to polaron formation and lattice-mediated attractive interactions that are not present in conventional semiconductors. Combined with near-unity site occupations at high doping levels, configurational entropy becomes central to the carrier distribution. This regime lies outside the scope of traditional semiconductor theory, which takes the weakly interacting electron gas as its foundation.

Herein, we take the carrier gas picture at face value. Using statistical mechanics, we treat the population of charge carriers within the grand canonical ensemble of a lattice gas, which naturally accommodates collective and critical phenomena. Two driving forces, the energetic interactions between carriers and configurational entropy, compete in controlling the degree of carrier aggregation. The emergence of phase-separated domains and hysteretic bistability can then be understood in analogy to a collective vapor--liquid phase transition. We proceed as follows: We first review the lattice gas in terms of its microscopic states and interactions, before exploring its equilibrium and dynamical behavior through Monte Carlo simulations. Using mean-field analysis, we formalize the grand canonical probability distribution and close by connecting this statistical picture to experimental observables in transistor operation.

\section{Main}
\subsection{The lattice gas model}

Distributing charge carriers across the available sites in an OMC gives rise to a vast number of possible configurations, known as microstates. Meanwhile, device operation probes macroscopic properties, such as the total channel conductivity or impedance, which represent ensemble averages over this microscopic landscape. This separation of scales motivates a statistical treatment of the system's equilibrium properties. We focus on the subsystem responsible for the drain current: localized electronic charge carriers strongly coupled to local lattice distortions. This subsystem can be effectively decoupled from the bulk material by a separation of timescales between mobile carriers and large-scale structural rearrangements. We thus treat the molecular backbone as a quasi-static grid of sites hosting the carrier population. Coarse-graining into a lattice, each site represents a mesoscopic region of the polymer, for which we here consider a binary occupation encoded by $n_i \in \{0,1\}$. This picture maps onto a lattice-gas model, mathematically equivalent to the Ising model, which we use as a deliberately simple toy framework to isolate and study the role of carrier--carrier interactions.

Charge carriers in conjugated polymers are not free electrons but quasi-particles often referred to as polarons, where the electronic charge is strongly coupled to local, fast vibrations of the molecular lattice \cite{ghosh2020excitons}. Rather than being point-like, the electronic charge is delocalized over several repeat units (intra- and interchain), which reduces the Coulomb self-energy and can lead to sizable dipole moments and polarizability. In coupling to the lattice relaxation, it deepens the associated potential well and, for sufficiently strong electron–phonon coupling, can make it energetically favorable for a second carrier to occupy the same region. As estimated in Appendix~\ref{app:estimate_coupling}, these cooperative, short-range effects can outweigh the residual like-charge repulsion, motivating our central assumption of a net effective attraction between charge carriers. This energetic stabilization is analogous, in spirit, to the attractive part of intermolecular potentials (e.g., Lennard--Jones) that drives the condensation of classical fluids from a vapor. We stress that the heterogeneous environment in OMCs implies that polaronic carriers are not identical but exhibit a distribution in spatial extent, energy, and coupling strength. This notion may be captured by the concept of `weighted polarons,' which we neglect here for simplicity.

In operation, an OMC is not an isolated system; it is in contact with a thermal environment (the `heat bath') and a reservoir of dopant ions (the electrolyte). By the separation of timescales, the number of molecular host sites constitutes a constant effective volume ($V$) for the electronic subsystem. The exchange of energy with the heat bath is governed by the temperature ($T$), while the exchange of particles (charge carriers) with the reservoir is governed by the chemical potential ($\mu$). This perspective, allowing for particle exchange within a fixed effective volume, naturally leads to the grand canonical ensemble. For a microstate $X$ with $N(X)=\sum_i n_i$ carriers, the probability follows the Boltzmann distribution:
\begin{equation}
\label{eq:boltzmann_distribution}
    \mathbb{P}[X]\propto \exp\left(-\frac{E(X)-\mu N(X)}{k_\mathrm{B}T}\right),
\end{equation}
where $E(X)$ is the internal energy of the microstate and $N(X) = \sum_i n_i$ is the total number of carriers. For simplicity, we set $k_\mathrm{B}=1$ henceforth. $E(X)$ is defined by the Hamiltonian for our lattice gas:
\begin{equation}
     H[X] = -J_0 \sum_{\langle i,j \rangle} n_i n_j,
    \label{eq:hamiltonian}
\end{equation}
where $J_0 > 0$ is the microscopic attractive interaction energy between nearest-neighbor carriers motivated previously. We interpret $\mu$ as a fixed reservoir parameter (control parameter) and will later connect it to the electrochemical potential in device operation.

While the probability of a single microstate may be known, the probability of an observable macrostate (e.g., one with a specific carrier density $\rho=N/V$) must also account for its degeneracy, that is, the total number of microstates, $\Omega$, that may realize it. The probability of such a macrostate is proportional to the Boltzmann factor weighted by this degeneracy: $\mathbb{P}(\text{macrostate}) \propto \Omega \cdot e^{-(E-\mu N)/T}$. This degeneracy is quantified by the configurational entropy, $S = \ln \Omega$. Substituting into the probability expression reveals the central role of the thermodynamic potential:
\begin{equation}
    \mathbb{P}(\text{macrostate}) \propto e^S e^{-(E-\mu N)/T} = e^{-\Phi/T},
\end{equation}
where the grand potential is defined as $\Phi = E - TS - \mu N = F - \mu N$, with $F = E - TS$ being the Helmholtz free energy. The system's equilibrium state is the one that maximizes this probability, i.e., which minimizes $\Phi$. This setup represents the fundamental thermodynamic balance between the energetic drive for `organization' (minimizing $E$), the entropic drive for `chaos' (maximizing $S$), and the coupling to the particle reservoir ($-\mu N$).

\subsection{Monte Carlo simulation}

We investigate this model using Markov chain Monte Carlo (MCMC) simulations based on Metropolis acceptance \cite{metropolis1953equation}, sampling from the grand canonical ensemble. The algorithm (Alg.~\ref{alg:gcmc}) repeatedly attempts to change the state of a randomly selected lattice site. A proposed flip (e.g., creating a carrier at an empty site) results in a change in the system's energy and particle number, $\Delta E$ and $\Delta N$. The move is accepted with probability $P_{\text{accept}} = \min(1, e^{-(\Delta E - \mu \Delta N) / T})$, thus satisfying detailed balance with respect to the grand canonical distribution and ensuring that the simulation converges to the system's equilibrium state at fixed $\mu$ and $T$. We provide a simulation tool under \cite{bongartz2025rustlatticesimulator}.

\begin{algorithm}[t]
\caption{Grand-canonical MCMC}
\label{alg:gcmc}
\begin{algorithmic}[1]
\REQUIRE $L$, $T$, $\mu$, $J_0$, $N_\mathrm{sw}$
\STATE Initialize $\{n_i\}$ with $n_i \sim \mathrm{Bernoulli}(\rho_0)$
\FOR{$s = 1$ \textbf{to} $N_\mathrm{sw}$}
    \FOR{$k = 1$ \textbf{to} $L^2$}
        \STATE Pick site $i$ uniformly at random
        \STATE $h_i \leftarrow \sum_{j \in \mathcal{N}(i)} n_j$ \COMMENT{neighbor count}
        \STATE $\Delta N \leftarrow 1 - 2n_i$
        \STATE $\Delta E \leftarrow -J_0 \, \Delta N \, h_i$
        \STATE $\Delta H \leftarrow \Delta E - \mu \, \Delta N$ \COMMENT{$\Delta(E-\mu N)$}
        \STATE $P_\mathrm{acc} \leftarrow \min(1, \, e^{-\Delta H / T})$
        \IF{$U \sim \mathrm{Uniform}(0,1) < P_\mathrm{acc}$}
            \STATE $n_i \leftarrow 1 - n_i$
        \ENDIF
    \ENDFOR
\ENDFOR
\end{algorithmic}
\end{algorithm}

\begin{figure}[b]
\includegraphics[width=\columnwidth]{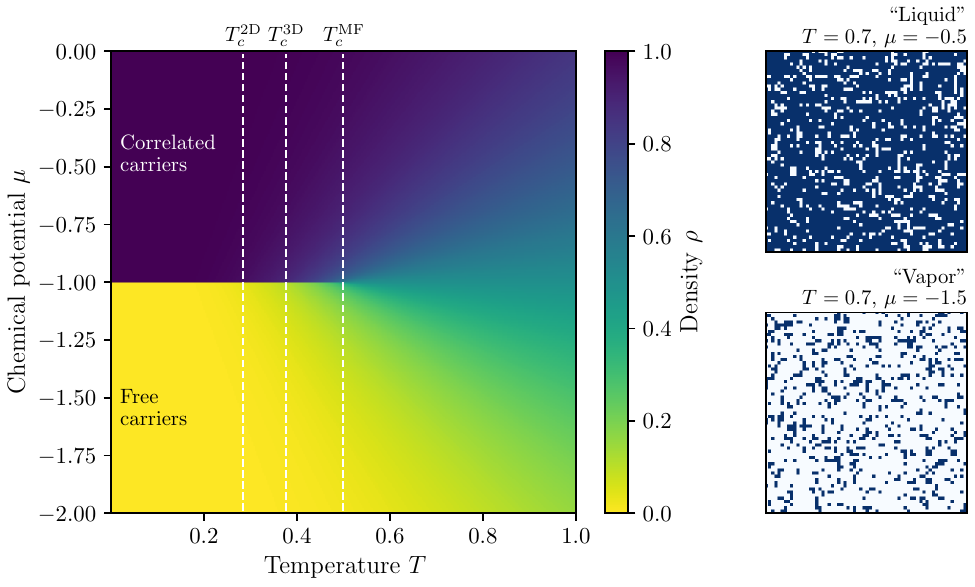}
\caption{\label{fig:phase_diagram} Phase diagram of the lattice gas model obtained from the mean-field grand potential (Sec.~\ref{subsec:meanfield}) as a function of temperature $T$ and chemical potential $\mu$ ($J = 1.0$). At high temperatures, $\rho$ varies smoothly with $\mu$ (supercritical regime); below the critical temperature, a first-order transition occurs at $\mu_c = -J$, separating a low-density ``vapor'' phase (free carriers) from a high-density ``liquid'' phase (correlated carriers). Dashed lines indicate the mean-field critical temperature $T_c^\mathrm{MF} = J/2 = 0.5$ and, for comparison, the exact critical temperatures of the 2D square-lattice Ising model ($T_c^\mathrm{2D} \approx 0.284$) \cite{onsager1944crystal} and the 3D simple cubic Ising model ($T_c^\mathrm{3D} \approx 0.376$) \cite{ferrenberg1991critical}. Side panels show Monte Carlo snapshots of the 2D lattice at the indicated $(T,\mu)$ values.}
\end{figure}

The equilibrium phase structure of this model is summarized in Fig.~\ref{fig:phase_diagram}, which shows the mean-field phase diagram obtained from the analysis of Sec.~\ref{subsec:meanfield}, with Monte Carlo snapshots at representative points. The diagram reveals two distinct regimes. At high temperatures, the carrier density changes smoothly and continuously with $\mu$, analogous to a supercritical fluid. At low temperatures, a discontinuous jump appears in the density at a critical chemical potential $\mu_c = -J$ (Appendix~\ref{app:meanfield}). This is a first-order phase transition, separating a low-density ``vapor'' phase of largely independent carriers from a high-density ``liquid'' phase of correlated, stabilized carriers. Near this phase transition line, the system exhibits metastability; if parameters are changed such that a new phase becomes favorable, the system can remain trapped in the previous macrostate for an extended period of time, as an energy barrier for nucleation of the new phase must be overcome. 

\begin{figure}[t]
\includegraphics[width=\columnwidth]{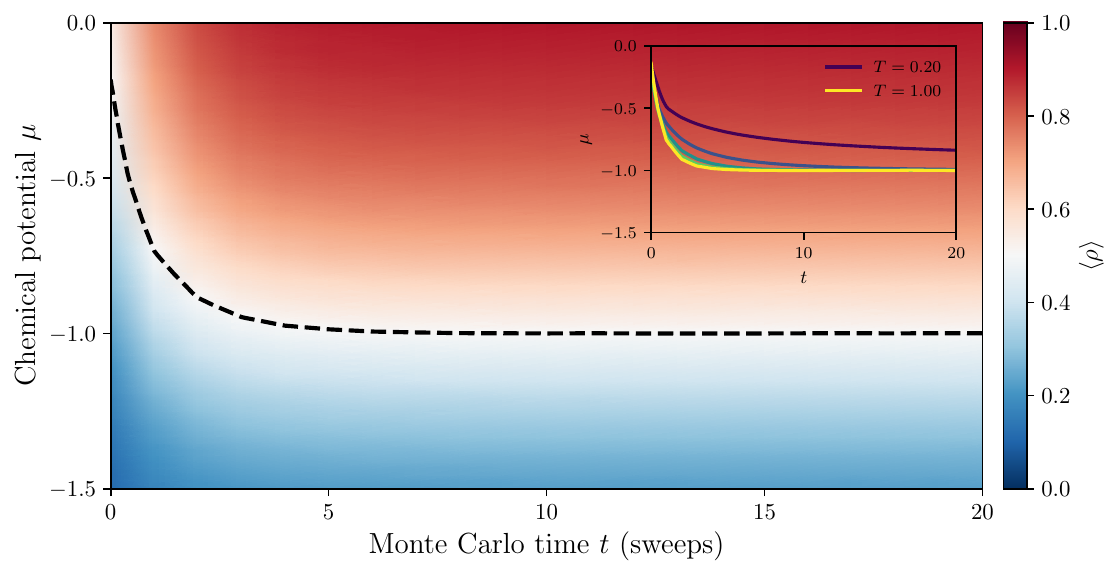}
\caption{\label{fig:dynamical_phase_mu} Dynamical phase diagram showing the average carrier density $\langle \rho \rangle$ as a function of Monte Carlo time $t$ and chemical potential $\mu$ at fixed temperature $T = 0.8$ 
($J = 1.0$, $\rho_0 = 0$, averaged over $10^3$ 
realizations). The dashed line marks $\langle \rho \rangle = 0.5$, indicating the dynamical phase boundary. Inset: the $\langle \rho \rangle = 0.5$ contour for temperatures $T = 0.2$ (blue) to $T = 1.0$ (yellow), showing that metastable states become increasingly long-lived at lower temperatures.}
\end{figure}

The simulations use a nearest-neighbor coupling $J_0 = 0.5$, corresponding to a mean-field coupling $J = z J_0/2 = 1.0$ for coordination number $z = 4$. The mean-field prediction $T_c^\mathrm{MF} = J/2 = 0.5$ overestimates the exact critical temperature, a well-known feature of mean field in low dimensions that reflects the neglect of fluctuations. For the 2D square-lattice Ising model, to which our lattice gas is equivalent, Onsager's exact result \cite{onsager1944crystal} gives $T_c^\mathrm{2D} = J_0/(2\ln(1+\sqrt{2})) \approx 0.284$, which our Monte Carlo implementation reproduces. The discrepancy between mean field and the exact result shrinks with increasing dimensionality: for the 3D simple cubic lattice \cite{ferrenberg1991critical} $T_c^\mathrm{3D} \approx 0.376$, already within a factor $\sim 1.33$ of the mean-field value. An order-of-magnitude mapping to physical temperature can be obtained from the coupling estimates in Appendix~\ref{app:estimate_coupling}: the net attractive interaction $|E_\mathrm{eff}| \approx 70$--$85\,\mathrm{meV}$ for polymer semiconductors suggests $J_0 \sim 3\,k_\mathrm{B}T$ at $T = 300\,\mathrm{K}$, placing room-temperature operation well below $T_c$ for strongly coupled OMC systems.

We show a dynamical phase diagram at fixed temperature in Fig.~\ref{fig:dynamical_phase_mu}. The system is initialized in an empty state ($\rho_0=0$), and independent simulations are performed at different values of the chemical potential $\mu$, each evolving at fixed thermodynamic parameters. The color map shows the average density $\langle \rho \rangle$ as a function of Monte Carlo time $t$ and chemical potential $\mu$. The dashed line marks the dynamical phase boundary at $\langle \rho \rangle = 0.5$. At short times, this boundary is displaced from the equilibrium value $\mu_c = -J$; the system remains in its initial low-density state for an extended period before crossing over to high density. As $t \to \infty$, the boundary converges to $\mu_c$, recovering the equilibrium prediction. The inset shows how this convergence slows as the temperature is lowered, with the lowest temperatures ($T \lesssim T_c$) exhibiting metastable states separated by a nucleation barrier.

We note that the specific shape of the dynamical boundary depends on the choice of sampling algorithm. This toy model uses Metropolis acceptance, where each Monte Carlo sweep can be interpreted as a Poisson process with transition rates set by the acceptance probability (line 9 of Alg.~\ref{alg:gcmc}). An alternative is Kawasaki dynamics, where pairs of neighboring sites exchange occupation rather than flip individually, conserving total particle number and mimicking diffusive hopping.

\subsection{Mean-field analysis}
\label{subsec:meanfield}

The regular square lattice in our simulation represents a significant idealization. Real OMCs are high-entropy materials whose long-chain polymers possess vast conformational degrees of freedom, resulting in substantial structural heterogeneity. While OMC microstructures can comprise mixtures of (semi-)crystalline and amorphous regions, a coarse-grained model can describe the collective system properties through the average interaction experienced by carriers, rather than through specific local geometric arrangements. To this end, we relax all geometrical constraints by adopting a mean-field approximation, effectively treating the system as a fully connected network. We note that this approximation is better justified for the disordered three-dimensional OMC films of practical interest than for the two-dimensional square lattice used in our simulations: mean field becomes exact in the limit of large coordination number, and real OMCs combine higher dimensionality with substantial structural disorder.

The interaction energy of a given state depends only on the total number of carriers, $N(X)$, not their specific arrangement. For a system with $V$ total sites, this leads to the coarse-grained interaction energy
\begin{equation}
    E(X) = - \frac{J}{V}\,N(X)^2,
\end{equation}
where $J$ is the effective mean-field coupling that encapsulates the underlying microscopic interactions (Appendix~\ref{app:meanfield}). Substituting this expression into the grand canonical Boltzmann distribution gives
\begin{equation}
    \mathbb{P}[X] \propto \exp\left(\frac{J}{TV}N(X)^2 + \frac{\mu}{T}N(X)\right).
\end{equation}
This formulation connects the probability of a single microstate to its macroscopic density $\rho(X) = N(X)/V$. To find the probability of observing a macrostate with a given density $\rho$, we must multiply by the number of ways to arrange $N=\rho V$ carriers on $V$ sites. This combinatorial factor gives rise to the system's entropy. A standard derivation using Stirling's approximation (Appendix~\ref{app:entropy}) allows us to write the probability distribution for the density $\rho$ as
\begin{equation}
    \mathbb{P}[\rho] \propto \exp\left(V \left[ \frac{J\rho^2 + \mu\rho}{T} - \left( \rho \ln \rho + (1-\rho)\ln(1-\rho) \right) \right] \right).
    \label{eq:prob_density}
\end{equation}
\begin{figure}[t]
\includegraphics[width=\columnwidth]{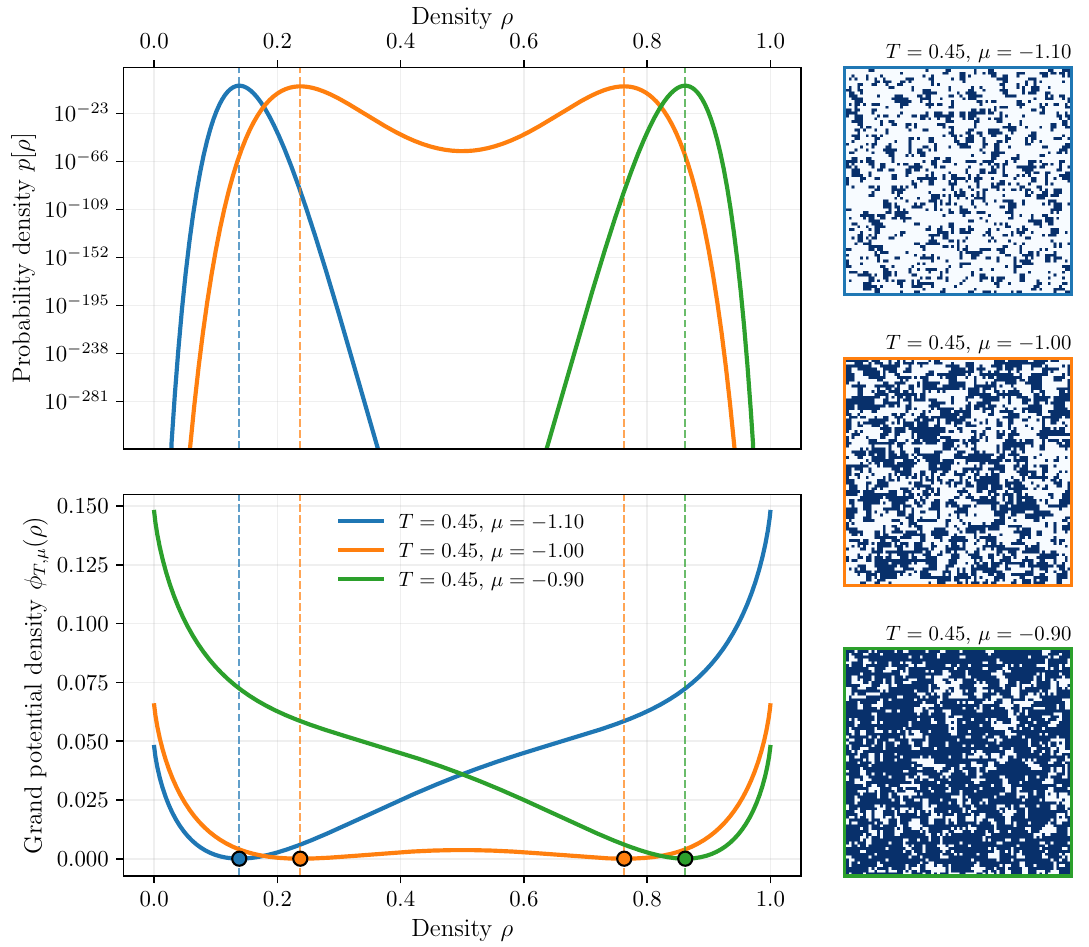}
\caption{\label{fig:prob_and_fe} Interplay of probability and grand potential in the mean-field model ($J=1.0$). (Top) Probability density $p[\rho]$ as a function of carrier density $\rho$. (Bottom) The corresponding grand potential landscape, $\phi_{T,\mu}(\rho)$ (shifted vertically for clarity). The peaks in probability correspond directly to the minima in the grand potential. As the chemical potential $\mu$ increases, the most probable state (global minimum) shifts from a low-density ``vapor'' phase (blue) to a high-density ``liquid'' phase (green), passing through a coexistence point where both phases are equally probable (orange). Insets show representative simulation snapshots for each regime.}
\end{figure}
For a large system ($V \gg 1$), the probability distribution is sharply peaked around the density $\rho^*$ that maximizes the term in the exponent. We can identify this exponent (per unit volume) as $-\phi_{T,\mu}(\rho)/T$, where 
\begin{equation}
\phi_{T,\mu}(\rho) = f(\rho) - \mu\rho
\end{equation}
is the mean-field grand potential density, with
\begin{equation}
    f(\rho) = -J\rho^2 + T\bigl[\rho\ln\rho + (1-\rho)\ln(1-\rho)\bigr]
\end{equation}
being the Helmholtz free energy density. The relationship between the probability and the grand potential landscape is visualized in Fig.~\ref{fig:prob_and_fe}. The peaks in the probability distribution $\mathbb{P}[\rho]$ correspond to minima in the grand potential density $\phi_{T,\mu}(\rho)$, which represent stable or metastable equilibrium states. At a temperature below the critical point, the landscape is non-convex and exhibits two minima corresponding to low- and high-density phases. The chemical potential controls the relative depth of these minima: for $\mu < \mu_c$, the low-density state is globally stable; for $\mu > \mu_c$, the high-density state is favored; at $\mu = \mu_c$, both minima have equal depth (coexistence). As $\mu$ crosses $\mu_c$, the global minimum of the grand potential jumps discontinuously from the low-density to the high-density state. This jump is the mathematical signature of a first-order phase transition.

\subsection{Connection to device operation}

In the lattice-gas description above, the charge carrier population is controlled by the chemical potential $\mu$ that appears in the grand-canonical Boltzmann weight of Eq.~\eqref{eq:boltzmann_distribution}. In the mean-field analysis, this $\mu$ enters Eq.~\eqref{eq:prob_density} as an abstract control parameter that tilts the grand-potential landscape $\phi_{T,\mu}(\rho)$ and thereby selects the most probable carrier density. Sweeping $\mu$ along the $\mu$-axis of the phase diagram is equivalent to moving the system across the vapor--liquid phase boundary.

In device operation, this control parameter is set by the electrochemical environment provided by the electrolyte and gate. The electrolyte acts as a reservoir of ions that can exchange charge with the polymer channel, while the gate electrode controls the electrostatic potential of this reservoir relative to the channel. The electrochemical potential of a charge carrier of charge $q$ is
\begin{equation}
    \mu_{\text{ec}} = \mu_{\text{chem}} + q \phi,
\end{equation}
where $\mu_{\text{chem}}$ is the chemical contribution and $\phi$ is the local electrostatic potential. The reservoir (electrolyte plus gate) is characterized by a fixed electrochemical potential, $\mu_{\text{reservoir}}$, and at equilibrium, the electrochemical potential in the channel must match that of the reservoir. Solving for the channel chemical potential that enters the lattice-gas weight gives
\begin{equation}
    \mu_{\text{eff}} = \mu_{\text{reservoir}} - q\phi_{\text{ch}}.
\end{equation}
Only a fraction of the applied gate voltage drops across the channel and couples to the hosting sites, which we capture by a coupling factor $0<\gamma\leq 1$. At a local channel potential $V_{\text{ch}}(x)$, we approximate the electrostatic potential as $\phi \approx \gamma\,(V_G - V_{\text{ch}})$, so that the effective chemical potential entering the grand-canonical weight becomes
\begin{equation}
    \mu_{\text{eff}}(V_G,V_{\text{ch}}) 
    = \mu_{\text{reservoir}} - \gamma\,q\bigl(V_G - V_{\text{ch}}\bigr),
    \label{eq:mu_eff_device}
\end{equation}
where $\gamma = 1$ corresponds to ideal coupling and $\gamma < 1$ accounts for partial screening and voltage division across gate, electrolyte, and channel. In the small-drain-bias limit where $V_{\text{ch}}\approx 0$, this reduces to the simpler form $\mu_{\text{eff}} \approx \mu_{\text{reservoir}} - \gamma q V_G$, making explicit that sweeping $V_G$ is equivalent to sweeping the control parameter $\mu$ in Eq.~\eqref{eq:prob_density}.

In the mean-field lattice gas, the occupation $\rho$ of a site with local effective chemical potential $\mu_{\text{eff}}$ obeys the self-consistency condition (Appendix~\ref{app:mean})
\begin{equation}
    \rho = \frac{1}{1 + \exp\!\left[-\frac{2J \rho + \mu_{\text{eff}}}{T}\right]},
    \label{eq:rho_mu_meanfield}
\end{equation}
where $J$ is again the effective mean-field interaction strength and $T$ is the dimensionless temperature. Combining Eqs.~\eqref{eq:mu_eff_device} and~\eqref{eq:rho_mu_meanfield} gives the local occupation $\rho(V_G,V_{\text{ch}})$ along the channel for a given gate and drain bias.

Equation~\eqref{eq:rho_mu_meanfield} determines the equilibrium occupation at fixed $\mu_{\text{eff}}$ but does not describe the system's transient response during a gate-voltage sweep. To capture this, we supplement the mean-field description with relaxational dynamics for the carrier density. In the grand canonical ensemble, carriers are exchanged with the electrolyte reservoir and their total number is not conserved, which motivates a non-conserved relaxation governed by the ordinary differential equation
\begin{equation}
  \tau \frac{\mathrm{d}\rho}{\mathrm{d}t} = -\frac{\partial \phi}{\partial \rho} = 2J\rho - T\ln\!\left(\frac{\rho}{1-\rho}\right) + \mu_{\text{eff}},
     \label{eq:model_a}
 \end{equation}
where $\tau$ is a phenomenological relaxation timescale. At steady state ($\mathrm{d}\rho/\mathrm{d}t=0$), this recovers the self-consistency condition Eq.~\eqref{eq:rho_mu_meanfield}. Equation~\eqref{eq:model_a} is the deterministic, mean-field counterpart of the stochastic Monte Carlo dynamics illustrated in Fig.~\ref{fig:dynamical_phase_mu}: both describe non-conserved relaxation of the carrier density, with metastability escape governed by the spinodal limit in the deterministic case and by thermal nucleation in the stochastic case.

To connect this statistical picture to the drain current, we consider a channel of width $W$, thickness $t$, and length $L$ biased by a drain voltage $V_D$. The drain voltage imposes a potential drop $V_{\text{ch}}(x)\in[0,V_D]$ along the channel, so that $\mu_{\text{eff}}$ and $\rho$ become position dependent. With carrier mobility $\mu_\mathrm{tr}$ and a density of available sites $n_{\text{max}}$, the local conductivity is
\begin{equation}
    \sigma\bigl(V_G,V_{\text{ch}}\bigr)
    = q\,\mu_\mathrm{tr}\,n_{\text{max}}\,\rho\bigl(V_G,V_{\text{ch}}\bigr),
\end{equation}
where the linear dependence on $\rho$ is a mean-field approximation that neglects spatial correlations such as percolation. Its validity is examined numerically for the 2D lattice used in our simulations in Appendix~\ref{app:transport_check}. 
\begin{figure}[b]
\includegraphics[width=\columnwidth]{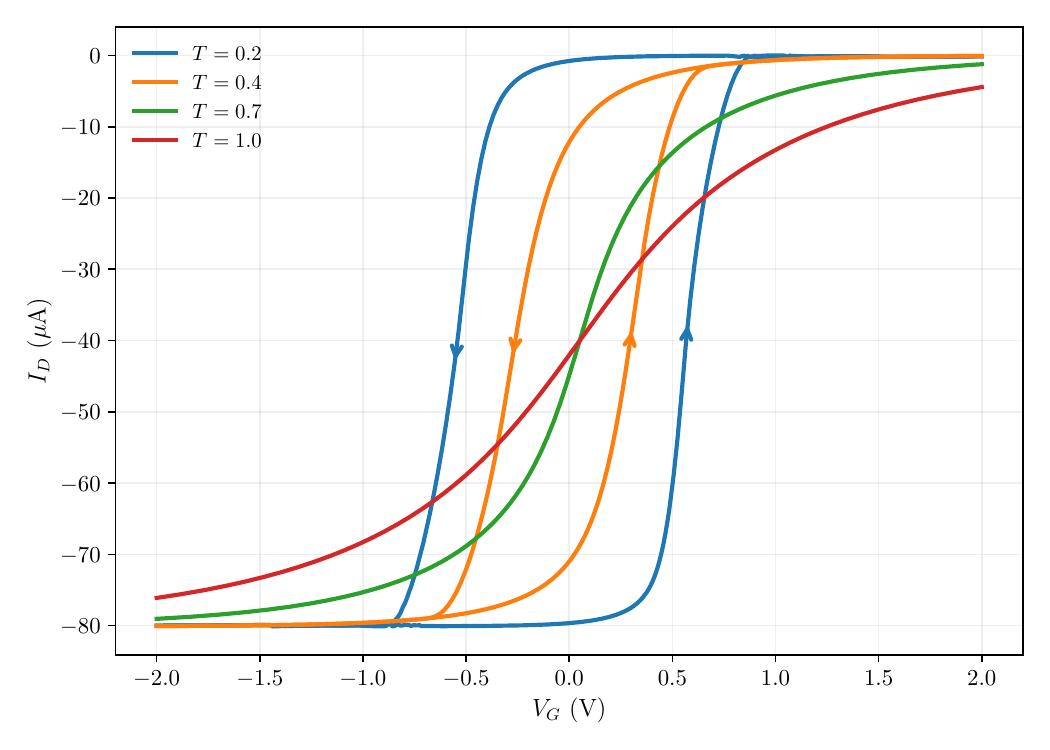}
\caption{\label{fig:transfer_characteristics}
Predicted transfer characteristics from the lattice-gas model: drain current $I_D$ versus gate voltage $V_G$ at $V_D = 0.1\,\mathrm{V}$ for several reduced temperatures $T$. Above the critical temperature ($T=0.7,1.0$), the response is smooth; below ($T=0.2,0.4$), the finite-rate gate-voltage sweep produces hysteresis loops (arrows indicate sweep direction). The loops are computed from Eq.~\eqref{eq:model_a} by sweeping $V_G$ forward and backward and evaluating the drain current via Eq.~\eqref{eq:Id_general} at each step. Parameters: $J=1.0$, $\mu_{\mathrm{reservoir}}=-1.0$, $\gamma=1.0$, $W=50\,\si{\micro\metre}$, $L=100\,\si{\micro\metre}$, $t=100\,\mathrm{nm}$, $n_{\max}=1\times 10^{21}\,\mathrm{cm}^{-3}$, $\mu_{\mathrm{tr}}=1\,\mathrm{cm}^2\mathrm{V}^{-1}\mathrm{s}^{-1}$.}
\end{figure}

In steady state, the drain current $I_D$ is constant along the channel. Neglecting diffusion and considering only drift transport, the current obeys
\begin{equation}
	I_D = -\sigma(V_G,V_{\text{ch}})\,W t\,\frac{\mathrm{d}V_{\text{ch}}}{\mathrm{d}x}.
\end{equation}
Eliminating the spatial coordinate $x$ in favour of the channel potential $V_{\text{ch}}\in[0,V_D]$ and integrating yields
\begin{equation}
    I_D(V_G,V_D)
    = -G_0 \int_{0}^{V_D} \rho\bigl(V_G,V_{\text{ch}}\bigr)\,\mathrm{d}V_{\text{ch}},
    \label{eq:Id_general}
\end{equation}
where
\begin{equation}
    G_0 = \frac{q\,\mu_\mathrm{tr}\,n_{\text{max}}\,W t}{L}
\end{equation}
collects the geometric and kinetic prefactors (Appendix~\ref{app:drift}). Together, Eqs.~\eqref{eq:prob_density}, \eqref{eq:mu_eff_device}, \eqref{eq:model_a}, and~\eqref{eq:Id_general} provide a link between the grand-canonical description and the measured drain current as a function of gate and drain bias. In the non-interacting limit ($J=0$), Eq.~\eqref{eq:Id_general} reduces to a closed expression for $I_D(V_G,V_D)$ obtained by analytic integration of the logistic occupation (Appendix~\ref{app:interaction_limits}). In the small-drain limit, $V_D\to 0$, the channel potential is nearly uniform and $\rho(V_G,V_{\text{ch}})\approx\rho(V_G,0)$, so that Eq.~\eqref{eq:Id_general} simplifies to
\begin{equation}
    I_D(V_G,V_D) \simeq -G_0\,\rho(V_G,0)\,V_D.
\end{equation}
Figure~\ref{fig:transfer_characteristics} shows the resulting drain current $I_D$ as a function of $V_G$ at fixed $V_D$, assuming typical material and device parameters. Above the critical temperature, the response is smooth and reversible, whereas below, the first-order phase transition gives rise to hysteretic switching. When the gate voltage is swept at a finite rate, the relaxation dynamics of Eq.~\eqref{eq:model_a} keep the system on its current metastable branch until the spinodal is reached and the local minimum of $\phi(\rho)$ vanishes, at which point it transitions to the opposite branch. This produces distinct forward and backward traces. The hysteresis width also depends on the ratio of sweep rate to relaxation rate $\tau$: faster sweeps produce wider loops, consistent with sweep-rate-dependent hysteresis commonly observed in OECTs \cite{bisquert2023hysteresis}.

We note that the fundamental control parameter for the metastability is not temperature alone, but the ratio $J/T$, which encapsulates the balance between energetic and entropic contributions. This is consistent with experimental reports. For example, Ji et al. reported that OECTs treated with a hydrophobic agent exhibit an increased hysteresis width. In our framework, this treatment enhances coupling by repelling the aqueous solvent, thereby increasing the effective $J$ \cite{ji2021mimicking}. Complementary to this, Bongartz et al. reported the inverse trend of shrinking hysteresis with rising temperature \cite{bongartz2024bistable}. In general, these phenomena are characteristic of OECT systems engineered to be in a `strong-coupling' regime, such as those employing ionic liquids and modified OMC composition, where the ratio $J/T$ is significant.

\section{Conclusion}
We have presented a minimal statistical-mechanics framework based on a lattice gas that describes the quasi-steady state of OMCs in the high charge density regime. Our results demonstrate that phenomena such as phase separation and hysteretic bistability can be understood in analogy to a collective vapor--liquid phase transition, driven by the competition between interaction energy and configurational entropy. The direct connection between gate-voltage sweeps and trajectories through the thermodynamic phase diagram provides a natural interpretation of the non-trivial transfer characteristics observed in devices operating in a strong-coupling regime. While this model is a strong simplification of the intricate molecular and electronic landscape of real OMCs, its ability to capture the essential features of a phase transition illustrates the principle of universality: near a critical point, the qualitative macroscopic behavior is governed chiefly by symmetries, dimensionality, and conservation laws rather than microscopic detail. 

We close by noting that our mean-field treatment assumes local equilibrium at each point in the channel. A more complete description of spatial inhomogeneities, such as the nanoscale domains observed by Wu et al.~\cite{wu2024bridging}, would require a Landau--Ginzburg formulation with a gradient term penalizing spatial variations in density. The resulting Cahn--Hilliard dynamics would describe domain coarsening and interface evolution, building naturally on the free energy landscape established here. More broadly, our work suggests that refined descriptions incorporating drift-diffusion transport and explicit spatial structure can be built on the same statistical-mechanical backbone, with $J$ serving as a mesoscopic parameter encapsulating the net effect of underlying microscopic interactions. We aim to pursue these extensions in future work.

\begin{acknowledgments}
The author thanks K. Leo, H. Kleemann, A. Hofacker, and P. Zschoppe for helpful remarks.
\end{acknowledgments}

\appendix

\section{Estimate of correlation energies}
\label{app:estimate_coupling}
We perform a simple estimate of the charge carrier interactions in inorganic (silicon) and carbon-based polymer semiconductors. We consider the systems in terms of their three-dimensional charge carrier density $n$ with the average carrier spacing as
\begin{equation}
    r \approx n^{-\frac{1}{3}}.
\end{equation}

Charge carriers in OMCs are not point-like but possess molecular and polaronic character, residing within an ionic environment. We attempt a rough estimate of the effective carrier--carrier interaction by considering (screened) Coulomb repulsion ($E_\mathrm{C}$) on the one hand, and stabilization through dipole-dipole ($E_\mathrm{dd}$), induced dipole ($E_\mathrm{ind}$), and lattice relaxation ($E_\mathrm{pol}$) interactions on the other, following
\begin{align}
    E_\mathrm{C} &= \frac{e^2}{4\pi\epsilon_0\epsilon_r} \cdot \frac{1}{r} \cdot \exp\left(-\frac{r}{\lambda_\mathrm{D}}\right), \label{eq:E_C} \\
    E_\mathrm{dd} &= -\frac{2p^2}{4\pi\epsilon_0\epsilon_r} \cdot \frac{1}{r^3}, \label{eq:E_dd} \\
    E_\mathrm{ind} &= - \frac{\alpha e^2}{8\pi\epsilon_0\epsilon_r^2} \cdot \frac{1}{r^4}, \quad \text{and} \label{eq:E_ind} \\
    E_\mathrm{pol} &= - \eta g^2 \hbar \omega_0, \label{eq:E_pol}
\end{align}
with $p$ the permanent dipole moment, $\alpha$ the polarizability volume, $g$ the dimensionless electron--phonon coupling constant in the Holstein model, and $\hbar\omega_0$ the characteristic phonon energy. The factor $\eta \leq 1$ accounts for the partial overlap of lattice distortions when two polarons share a common relaxation field. In electrochemically doped systems, mobile ions screen the Coulomb repulsion between carriers. We model this through the screening factor in Eq.~\eqref{eq:E_C}, with the Debye length
\begin{equation}
    \lambda_\mathrm{D} = \sqrt{\frac{\epsilon_0 \epsilon_r k_\mathrm{B} T}{2 n_\mathrm{ion} e^2}},
    \label{eq:debye_length}
\end{equation}
where $n_\mathrm{ion}$ is the ion concentration in the material. The net effective interaction energy between two carriers is then
\begin{equation}
    E_\mathrm{eff} = E_\mathrm{C} + E_\mathrm{dd} + E_\mathrm{ind} + E_\mathrm{pol},
    \label{eq:E_eff}
\end{equation}
with $E_\mathrm{eff} > 0$ corresponding to net repulsion and $E_\mathrm{eff} < 0$ to net attraction. The dimensionless correlation parameter
\begin{equation}
    \Gamma = \frac{E_\mathrm{eff}}{k_\mathrm{B} T}
    \label{eq:Gamma}
\end{equation}
quantifies the strength of carrier--carrier interactions relative to thermal fluctuations. For $|\Gamma| \gg 1$, correlations dominate and collective phenomena such as phase transitions become relevant. 

\begin{table*}
\caption{\label{tab:estimate_int_energies} Order-of-magnitude estimate of charge carrier interactions in silicon and polymer semiconductors ($T=300\,\text{K}$).
}
\begin{ruledtabular}
\begin{tabular}{l d{-1} d{-1} d{-1} d{-1} d{-1} d{-1} d{-1}}
\multicolumn{1}{r}{$n\,[\mathrm{cm}^{-3}]$} &
\multicolumn{1}{c}{$r\,[\mathrm{nm}]$} &
\multicolumn{1}{c}{$E_\mathrm{C}\,[\mathrm{meV}]$} &
\multicolumn{1}{c}{$E_\mathrm{dd}\,[\mathrm{meV}]$} &
\multicolumn{1}{c}{$E_\mathrm{ind}\,[\mathrm{meV}]$} &
\multicolumn{1}{c}{$E_\mathrm{pol}\,[\mathrm{meV}]$} &
\multicolumn{1}{c}{$E_\mathrm{eff}\,[\mathrm{meV}]$} &
\multicolumn{1}{c}{$\Gamma$} \\
\colrule
\\[-1ex]
\multicolumn{8}{l}{Si ($\epsilon^\text{Si}_r \approx 12$, $\hbar\omega_0=60\,\text{meV}$, $p=0\,\text{D}$, $\alpha=5\,\si{\cubic\angstrom}$, $g=0.1$, $\eta = 1.0$)}\\
\colrule
\\[-2ex]
$10^{16}$ & 46.4 & 2.59 & 0.00& 0.00 & -0.6 & 1.99 & 0.08 \\
$10^{18}$ & 10.0 & 12.00 & 0.00& 0.00 & -0.6 & 11.40 & 0.44 \\
$10^{19}$ & 4.6  & 25.85 & 0.00& 0.00 & -0.6 & 25.25 & 0.98 \\[0.5ex]
\multicolumn{8}{l}{Polymer ($\epsilon^\text{poly}_r \approx 3$, $\hbar\omega_0=180\,\text{meV}$, $p=3\,\text{D}$, $\alpha=100\,\si{\cubic\angstrom}$, $g=0.9$, $\eta = 0.5$, $n_\mathrm{ion} = n$)}\\
\colrule
\\[-2ex]
$10^{18}$ & 10.0 & 0.05 & 0.00 & 0.00 & -72.9 & -72.85 & -2.82 \\
$10^{20}$ & 2.2  & 0.00 & -0.37 & -0.37 & -72.9 & -73.65 & -2.85 \\
$10^{21}$ & 1.0  & 0.00 & -3.75 & -8.00 & -72.9 & -84.65 & -3.27 \\
\end{tabular}
\end{ruledtabular}
\end{table*}

Tab.~\ref{tab:estimate_int_energies} compares the estimates for both systems using archetypal material parameters. Silicon devices typically feature carrier densities of $n^\text{Si}\sim 10^{16}-10^{19}\,\mathrm{cm}^{-3}$ (moderate to high doping) with $\epsilon^\text{Si}_r \approx 12$. Note that in this simple picture, restricting the silicon system to its sheet density in a 2D inversion layer hardly changes the estimated numbers, since carrier spacing remains essentially the same. OMCs, on the other hand, operate in the range of $n^\text{poly}\sim 10^{18}-10^{21}\,\mathrm{cm}^{-3}$ with a permittivity of $\epsilon^\text{poly}_r \approx 3$ for the `dry' polymer backbone where carriers reside. The estimation suggests that carrier--carrier correlations are significantly stronger in polymer semiconductors than in silicon, and crucially, can be net attractive in nature. This conclusion holds even under strong screening by the electrolytic environment ($\epsilon^\text{poly}_r \approx 10$), which changes $\Gamma$ only marginally. Recently reported operando X-ray photon correlation spectroscopy measurements on a benchmark OMC have revealed mesoscale phase separation and domain coarsening dynamics consistent with net attractive carrier interactions \cite{wu2024bridging}, providing experimental support for this picture.

\section{Mean-field interaction and coexistence chemical potential}
\label{app:meanfield}

\subsection{Mean-field interaction energy}

Starting from the lattice-gas Hamiltonian of the main text,
\begin{equation}
    H[X] = -J_0 \sum_{\langle i,j\rangle} n_i n_j,
\end{equation}
we approximate the interaction energy in terms of the total number of
occupied sites $N(X)=\sum_i n_i$.

In the mean-field picture every site is taken to see the average occupation
$\rho = N/V$ of all other sites, so that the interaction term for a given
site $i$ is approximated as
\begin{equation}
-J_0 \sum_{j\in\mathrm{n.n.}(i)} n_i n_j
    \;\approx\; -\frac{J_0}{2} z\,\rho\, n_i,
\end{equation}
where $z$ is the coordination number. Summing over all sites gives
\begin{align}
    E(X) &\approx -\frac{J_0 z}{2} \,\rho \sum_i n_i \\
&= -\frac{J_0 z}{2}\,\frac{N(X)}{V}\,N(X) \\
&= -\frac{J_0 z}{2V}\,N(X)^2.
\end{align}
It is convenient to define a coarse-grained mean-field coupling
\begin{equation}
    J \equiv \frac{z}{2}\,J_0,
\end{equation}
so that the interaction energy can be written as
\begin{equation}
    E(X) = -\frac{J}{V}\,N(X)^2.
\end{equation}
Inserting this expression into the grand-canonical weight
$\mathbb{P}[X] \propto \exp[-(E(X)-\mu N(X))/T]$ used in the main text then
yields
\begin{equation}
    \mathbb{P}[X] \propto
\exp\!\left[
    \frac{J}{T V}N(X)^2 + \frac{\mu}{T}N(X)
\right].
\end{equation}

\subsection{Mean-field coexistence chemical potential}

In this mean-field description, the grand-potential density at temperature
$T$ and chemical potential $\mu$ can be written as
\begin{equation}
    \phi_{T,\mu}(\rho)
    = -J\rho^2
      + T\bigl[\rho\ln\rho + (1-\rho)\ln(1-\rho)\bigr]
      - \mu \rho,
    \label{eq:app_phi_meanfield}
\end{equation}
where $\rho$ is the carrier density ($0<\rho<1$) and $J>0$ is the effective
mean-field attraction defined above.

The coexistence (``critical'') chemical potential $\mu_c$ is obtained by
exploiting the particle--hole symmetry of the lattice gas. At coexistence, the low- and high-density phases are equally stable, which in mean field corresponds to a symmetry of the grand potential under 
$\rho \leftrightarrow 1-\rho$:
\begin{equation}
    \phi_{T,\mu_c}(\rho) = \phi_{T,\mu_c}(1-\rho)
    \qquad\text{for all } \rho.
\end{equation}
We therefore consider the difference
\begin{equation}
    \Delta\phi(\rho)
    = \phi_{T,\mu}(\rho) - \phi_{T,\mu}(1-\rho).
\end{equation}
Using Eq.~\eqref{eq:app_phi_meanfield}, we obtain
\begin{align}
    \Delta\phi(\rho)
    &= \Bigl[-J\rho^2 - \mu\rho\Bigr]
     - \Bigl[-J(1-\rho)^2 - \mu(1-\rho)\Bigr] \nonumber\\
    &= -J\rho^2 + J(1-2\rho+\rho^2)
       - \mu\rho + \mu(1-\rho) \nonumber\\
    &= J(1-2\rho) + \mu(1-2\rho) \nonumber\\
    &= (1-2\rho)\,(J+\mu).
\end{align}
The entropic contribution cancels in the difference because
\(\rho\ln\rho + (1-\rho)\ln(1-\rho)\) is symmetric under
$\rho \leftrightarrow 1-\rho$. At coexistence, we require
\(\Delta\phi(\rho)=0\) for all $\rho$, which is only possible if the prefactor
of $(1-2\rho)$ vanishes. This yields the mean-field coexistence chemical
potential
\begin{equation}
    \mu_c = -J.
\end{equation}
We note that this result, derived here from the mean-field grand potential, is in fact exact: it follows from the particle--hole symmetry of the lattice-gas Hamiltonian under $n_i \leftrightarrow 1 - n_i$ combined with $\mu \leftrightarrow -zJ_0 - \mu$, which holds at the microscopic level and is independent of dimensionality or the approximation scheme.

\section{Derivation of the mean-field entropy}
\label{app:entropy}

The entropy of a macrostate with carrier density $\rho$ arises from the number of ways to arrange $N = \rho V$ indistinguishable carriers on $V$ distinguishable sites, given by the binomial coefficient $\Omega = \binom{V}{N}$. Applying Stirling's approximation $\ln(n!) \approx n\ln n - n$ and expressing $N = \rho V$ yields, to leading order in $V$,
\begin{equation}
    S = \ln \Omega \approx -V\bigl(\rho\ln\rho + (1-\rho)\ln(1-\rho)\bigr),
\end{equation}
which defines the entropy density $s = S/V = -[\rho\ln\rho + (1-\rho)\ln(1-\rho)]$. This is the standard entropy of mixing on a lattice \cite{chandler1987introduction} and forms the entropic contribution to the mean-field free energy density in Eq.~\eqref{eq:prob_density}. Sub-leading corrections of order $\mathcal{O}(\ln V)$ are negligible in the thermodynamic limit.

\section{Mean-field lattice gas and local occupation}
\label{app:mean}

In the mean-field lattice-gas description, a site can be either undoped ($n=0$) or doped ($n=1$). For a configuration $X$ with $N(X)$ doped sites, the coarse-grained interaction energy is approximated as
\begin{equation}
    E(X) = - \frac{J}{V}\,N(X)^2 - \mu_{\text{eff}} N(X),
\end{equation}
where $J>0$ is the effective mean-field interaction, $V$ is the total number of sites, and $\mu_{\text{eff}}$ is the effective chemical potential.

Within mean field, each site feels the average occupation $\rho = \langle n_i\rangle = N/V$ of all other sites. The effective single-site internal energy can then be written as
\begin{equation}
    E_i(n_i) \approx - (2J \rho + \mu_{\text{eff}})\, n_i,
\end{equation}
so that the two possible energies are
\begin{equation}
    E_i(0) = 0, 
    \qquad
    E_i(1) = - (2J \rho + \mu_{\text{eff}}).
\end{equation}

The corresponding single-site grand-canonical partition function is
\begin{equation}
    Z_{\text{site}}
    = \sum_{n_i=0,1} \exp\!\left[-\frac{E_i(n_i)}{T}\right]
    = 1 + \exp\!\left[\frac{2J \rho + \mu_{\text{eff}}}{T}\right],
\end{equation}
and the average occupation $\rho = \langle n_i\rangle$ is
\begin{equation}
    \rho 
    = \frac{1}{Z_{\text{site}}}
      \sum_{n_i=0,1} n_i \,
      \exp\!\left[-\frac{E_i(n_i)}{T}\right]
    = \frac{\exp\!\left[\frac{2J \rho + \mu_{\text{eff}}}{T}\right]}
           {1 + \exp\!\left[\frac{2J \rho + \mu_{\text{eff}}}{T}\right]}.
\end{equation}
This can be written in the compact logistic form
\begin{equation}
    \rho = \frac{1}{1 + \exp\!\left[-\frac{2J \rho + \mu_{\text{eff}}}{T}\right]},
    \label{eq:app_rho_selfconsistent}
\end{equation}
which is the self-consistency condition quoted in Eq.~\eqref{eq:rho_mu_meanfield} of the main text. Together with the device coupling
\begin{equation}
    \mu_{\text{eff}}(V_G,V_{\text{ch}}) 
    = \mu_{\text{reservoir}} - \gamma\,q\bigl(V_G - V_{\text{ch}}\bigr),
    \label{eq:app_mu_eff}
\end{equation}
this defines the local occupation $\rho(V_G,V_{\text{ch}})$ along the channel.

\section{Validity of the mean-field transport approximation}
\label{app:transport_check}

The mean-field transport model assumes a local conductivity $\sigma \propto \rho$, neglecting spatial correlations such as percolation and current-path inhomogeneity. To assess this approximation, we compute the effective conductance of equilibrated 2D lattice-gas configurations by solving a resistor network in which each nearest-neighbor bond has conductance $g_{ij} = n_i n_j$, and compare the result to the mean-field prediction $G/G_{\mathrm{max}} = \rho$.

\begin{figure}[t]
\includegraphics[width=\columnwidth]{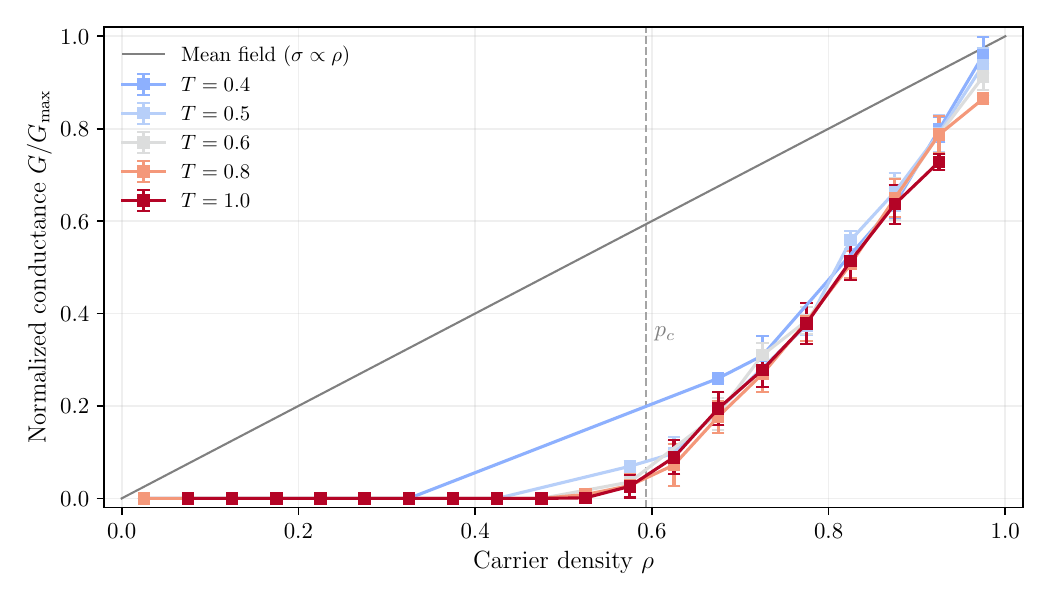}
\caption{\label{fig:conductance_verification}
Normalized effective conductance $G/G_{\mathrm{max}}$ of equilibrated 2D lattice-gas configurations as a function of carrier density $\rho$, compared to the mean-field prediction $G/G_{\mathrm{max}} = \rho$. Each point represents the conductance of a resistor network with bond conductances $g_{ij} = n_i n_j$, solved on Monte Carlo snapshots at the indicated temperatures ($J = 1.0$, $50\times50$ lattice). The dashed line marks the 2D site-percolation threshold $p_c \approx 0.59$ \cite{newman2000efficient}. Error bars denote the standard deviation over independent snapshots.}
\end{figure}

Figure~\ref{fig:conductance_verification} shows the normalized conductance as a function of carrier density for several temperatures. At intermediate densities, the mean-field prediction systematically overestimates the conductance: the 2D lattice exhibits a site-percolation threshold at $\rho_c \approx 0.59$ \cite{newman2000efficient}, below which no conducting path spans the system. Above $\rho_c$, the conductance rises steeply but remains suppressed relative to the mean-field prediction, approaching it only as $\rho \to 1$ when the lattice becomes fully occupied.

This discrepancy reflects the well-known limitation of mean-field transport in spatially heterogeneous systems. However, we note three points relevant to the present work. First, the mean-field transport assumption is self-consistent with the mean-field thermodynamics used throughout: both discard spatial correlations, and both would need to be extended simultaneously, e.g., via a Cahn--Hilliard formulation coupled to drift-diffusion transport. Second, the qualitative features of the transfer characteristics are governed by the thermodynamic landscape and require only that the conductivity is a monotonically increasing function of carrier density. In the sub-critical regime where the model predicts hysteresis, the first-order phase transition confines the carrier density to $\rho \approx 0$ or $\rho \approx 1$, being precisely the limits where $\sigma \propto \rho$ is accurate. Finally, the present 2D verification is conservative with respect to the three-dimensional OMC films of practical interest: in 3D the site-percolation threshold drops to $p_c \approx 0.31$ (simple cubic) \cite{wang2013bond} compared to $p_c \approx 0.59$ in 2D, so the density window in which $\sigma \propto \rho$ fails is substantially narrower. A quantitative treatment of transport at intermediate densities would nonetheless require percolation or effective-medium corrections, which go beyond the scope of this work.

\section{Drift transport and current continuity}
\label{app:drift}

We now derive the expression for the drain current. Consider a channel of width $W$, thickness $t$, and length $L$, with coordinate $x$ measured from source ($x=0$) to drain ($x=L$). The local channel potential is $V_{\text{ch}}(x)$, with boundary conditions $V_{\text{ch}}(0)=0$ and $V_{\text{ch}}(L)=V_D$. The local carrier density is
\begin{equation}
    n(x) = n_{\text{max}}\,\rho\bigl(V_G,V_{\text{ch}}(x)\bigr),
\end{equation}
where $n_{\text{max}}$ is the density of available sites. Assuming drift-dominated transport with mobility $\mu_\mathrm{tr}$, the local conductivity is
\begin{equation}
    \sigma(x)
    = q\,\mu_\mathrm{tr}\,n(x)
    = q\,\mu_\mathrm{tr}\,n_{\text{max}}\,\rho\bigl(V_G,V_{\text{ch}}(x)\bigr).
\end{equation}
The current density is $j(x) = \sigma(x) E(x)$ with electric field $E(x) = -\mathrm{d}V_{\text{ch}}/\mathrm{d}x$. The total drain current $I_D$ is constant along the channel and given by $I_D = j(x)\,W t$, so that
\begin{equation}
    I_D = -q\,\mu_\mathrm{tr}\,n_{\text{max}}\,\rho\bigl(V_G,V_{\text{ch}}(x)\bigr)\,W t\,
    \frac{\mathrm{d}V_{\text{ch}}}{\mathrm{d}x}.
\end{equation}
Rearranging yields
\begin{equation}
    \frac{\mathrm{d}V_{\text{ch}}}{\mathrm{d}x}
    = - \frac{I_D}{q\,\mu_\mathrm{tr}\,n_{\text{max}}\,W t}\,
      \frac{1}{\rho\bigl(V_G,V_{\text{ch}}(x)\bigr)}.
\end{equation}
Separating variables and integrating from source to drain,
\begin{equation}
    \int_{0}^{V_D} \rho\bigl(V_G,V_{\text{ch}}\bigr)\,\mathrm{d}V_{\text{ch}}
    = -\frac{L}{q\,\mu_\mathrm{tr}\,n_{\text{max}}\,W t}\, I_D,
\end{equation}
where we have changed the integration variable from $x$ to $V_{\text{ch}}$, using $V_{\text{ch}}(0)=0$ and $V_{\text{ch}}(L)=V_D$. This step assumes that $V_{\text{ch}}(x)$ varies continuously and monotonically between source and drain, and that the local carrier density can be written as a single-valued function $\rho(V_G, V_{\text{ch}})$ at fixed $V_G$ (local quasi-equilibrium). In the equilibrium formulation, these assumptions break down in regimes where the mean-field equation of state becomes multivalued, which would require explicit treatment of domain boundaries and spatial coexistence. We address this issue by the dynamical treatment of Eq.~\eqref{eq:model_a}: at each gate voltage, the relaxation dynamics select a unique branch, so that $\rho(V_G, V_{\text{ch}})$ remains single-valued along the channel for a given sweep direction. A full treatment of spatial coexistence within the channel would require a Cahn--Hilliard formulation as noted before. We further note that the linear dependence $\sigma \propto \rho$ is itself a mean-field approximation (Appendix~\ref{app:transport_check}). We thus write
\begin{equation}
    I_D(V_G,V_D)
    = -G_0 \int_{0}^{V_D} \rho\bigl(V_G,V_{\text{ch}}\bigr)\,\mathrm{d}V_{\text{ch}}
    \label{eq:app_Id_general}
\end{equation}
with
\begin{equation}
    G_0 = \frac{q\,\mu_\mathrm{tr}\,n_{\text{max}}\,W t}{L}
\end{equation}
as stated in Eq.~\eqref{eq:Id_general} of the main text.

\section{Non-interacting and interacting limits}
\label{app:interaction_limits}

In the non-interacting limit ($J = 0$), the mean-field self-consistency relation Eq.~\eqref{eq:app_rho_selfconsistent} reduces to a simple logistic dependence of the occupation $\rho$ on the effective chemical potential $\mu_{\text{eff}}(V_G, V_{\text{ch}})$ defined in Eq.~\eqref{eq:app_mu_eff}. Inserting this non-interacting $\rho(V_G, V_{\text{ch}})$ into the integral expression for the drain current, Eq.~\eqref{eq:app_Id_general}, yields a closed-form expression for $I_D(V_G, V_D)$. For $J > 0$, Eq.~\eqref{eq:app_rho_selfconsistent} becomes genuinely non-linear and can admit multiple solutions for $\rho$ at fixed $(T, \mu_{\text{eff}})$. The corresponding grand-potential landscape $\phi_{T,\mu}(\rho)$ develops multiple minima, reflecting the coexistence of low- and high-density phases and giving rise to metastability. In this interacting case, the static self-consistency condition alone does not uniquely specify $\rho(V_G, V_{\text{ch}})$, and selecting the physically realized branch requires a dynamical prescription. We use the relaxational dynamics introduced in Eq.~\eqref{eq:model_a}, which evolves $\rho$ along the gradient of the grand potential and tracks the metastable branch occupied by the system until it is destabilized at a spinodal. The drain current $I_D(V_G, V_D)$ in Eq.~\eqref{eq:app_Id_general} is then evaluated numerically using the dynamically selected $\rho(V_G, V_{\text{ch}})$. This is the regime relevant for the history-dependent behavior discussed in the main text.

\bibliographystyle{apsrev4-2}
\bibliography{references.bib}

@article{paulsen2020organic,
  title = {Organic mixed ionic--electronic conductors},
  author = {Paulsen, Bryan D. and Tybrandt, Klas and Stavrinidou, Eleni and Rivnay, Jonathan},
  journal = {Nature Materials},
  volume = {19},
  number = {1},
  pages = {13--26},
  year = {2020},
  publisher = {Nature Publishing Group UK London}
}

@article{gkoupidenis2024organic,
  title = {Organic mixed conductors for bioinspired electronics},
  author = {Gkoupidenis, Paschalis and Zhang, Y. and Kleemann, H. and Ling, H. and Santoro, F. and Fabiano, Simone and Salleo, A. and Van De Burgt, Y.},
  journal = {Nature Reviews Materials},
  volume = {9},
  number = {2},
  pages = {134--149},
  year = {2024},
  publisher = {Nature Publishing Group UK London}
}

@article{Rivnay2018,
  author = {Rivnay, J. and Inal, S. and Salleo, A. and Owens, R. M. and Berggren, M. and Malliaras, G. G.},
  title = {Organic electrochemical transistors},
  journal = {Nature Reviews Materials},
  volume = {3},
  pages = {17086},
  year = {2018}
}

@article{wang2024designing,
  title = {Designing organic mixed conductors for electrochemical transistor applications},
  author = {Wang, Yazhou and Wustoni, Shofarul and Surgailis, Jokubas and Zhong, Yizhou and Koklu, Anil and Inal, Sahika},
  journal = {Nature Reviews Materials},
  volume = {9},
  number = {4},
  pages = {249--265},
  year = {2024},
  publisher = {Nature Publishing Group UK London}
}

@article{frisbie2026charge,
  title={Charge transport physics of organic conductors at high carrier densities},
  author={Frisbie, C Daniel and Jacobs, Ian E and Ren, Xinglong and Sirringhaus, Henning},
  journal={Nature Reviews Materials},
  pages={1--16},
  year={2026},
  publisher={Nature Publishing Group UK London}
}

@article{he2022sub,
  title = {Sub-band filling, {M}ott-like transitions, and ion size effects in {C}$_{60}$ single crystal electric double layer transistors},
  author = {He, Tao and Frisbie, C. Daniel},
  journal = {ACS Nano},
  volume = {16},
  number = {3},
  pages = {4823--4830},
  year = {2022},
  publisher = {ACS Publications}
}

@article{tjhe2024non,
  title = {Non-equilibrium transport in polymer mixed ionic--electronic conductors at ultrahigh charge densities},
  author = {Tjhe, Dionisius H. L. and Ren, Xinglong and Jacobs, Ian E. and D'Avino, Gabriele and Mustafa, Tarig B. E. and Marsh, Thomas G. and Zhang, Lu and Fu, Yao and Mansour, Ahmed E. and Opitz, Andreas and others},
  journal = {Nature Materials},
  volume = {23},
  number = {12},
  pages = {1712--1719},
  year = {2024},
  publisher = {Nature Publishing Group UK London}
}

@article{fratini2020charge,
  title = {Charge transport in high-mobility conjugated polymers and molecular semiconductors},
  author = {Fratini, Simone and Nikolka, Mark and Salleo, Alberto and Schweicher, Guillaume and Sirringhaus, Henning},
  journal = {Nature Materials},
  volume = {19},
  number = {5},
  pages = {491--502},
  year = {2020},
  publisher = {Nature Publishing Group UK London}
}

@article{ihnatsenka2015understanding,
  title = {Understanding hopping transport and thermoelectric properties of conducting polymers},
  author = {Ihnatsenka, Siarhei and Crispin, Xavier and Zozoulenko, I. V.},
  journal = {Physical Review B},
  volume = {92},
  number = {3},
  pages = {035201},
  year = {2015},
  publisher = {APS}
}

@article{kim2012thermoelectric,
  title = {Thermoelectric model to characterize carrier transport in organic semiconductors},
  author = {Kim, Gunho and Pipe, Kevin P.},
  journal = {Physical Review B},
  volume = {86},
  number = {8},
  pages = {085208},
  year = {2012},
  publisher = {APS}
}

@article{warren2023molecular,
  title = {Molecular p-doping induced dielectric constant increase of polythiophene films determined by impedance spectroscopy},
  author = {Warren, Ross and Blom, Paul W. M. and Koch, Norbert},
  journal = {Applied Physics Letters},
  volume = {122},
  number = {15},
  pages = {152108},
  year = {2023},
  publisher = {AIP Publishing}
}

@article{ortstein2021band,
  title = {Band gap engineering in blended organic semiconductor films based on dielectric interactions},
  author = {Ortstein, Katrin and Hutsch, Sebastian and Hambsch, Mike and Tvingstedt, Kristofer and Wegner, Berthold and Benduhn, Johannes and Kublitski, Jonas and Schwarze, Martin and Schellhammer, Sebastian and Talnack, Felix and others},
  journal = {Nature Materials},
  volume = {20},
  number = {10},
  pages = {1407--1413},
  year = {2021},
  publisher = {Nature Publishing Group UK London}
}

@article{vukmirovic2010carrier,
  title = {Carrier hopping in disordered semiconducting polymers: {H}ow accurate is the {M}iller--{A}brahams model?},
  author = {Vukmirovi{\'c}, Nenad and Wang, Lin-Wang},
  journal = {Applied Physics Letters},
  volume = {97},
  number = {4},
  pages = {043305},
  year = {2010},
  publisher = {AIP Publishing}
}

@article{koopmans2021carrier,
  title = {Carrier--carrier {C}oulomb interactions reduce power factor in organic thermoelectrics},
  author = {Koopmans, Marten and Koster, L. Jan Anton},
  journal = {Applied Physics Letters},
  volume = {119},
  number = {14},
  pages = {143301},
  year = {2021},
  publisher = {AIP Publishing}
}

@article{spano2014h,
  title = {{H}- and {J}-aggregate behavior in polymeric semiconductors},
  author = {Spano, Frank C. and Silva, Carlos},
  journal = {Annual Review of Physical Chemistry},
  volume = {65},
  number = {1},
  pages = {477--500},
  year = {2014},
  publisher = {Annual Reviews}
}

@article{koopmans2020electrical,
  title = {Electrical conductivity of doped organic semiconductors limited by carrier--carrier interactions},
  author = {Koopmans, Marten and Leivisk{\"a}, Miina A. T. and Liu, Jian and Dong, Jingjin and Qiu, Li and Hummelen, Jan C. and Portale, Giuseppe and Heiber, Michael C. and Koster, L. Jan Anton},
  journal = {ACS Applied Materials \& Interfaces},
  volume = {12},
  number = {50},
  pages = {56222--56230},
  year = {2020},
  publisher = {ACS Publications}
}

@article{bongartz2025electron,
  author = {Bongartz, Lukas M. and LeCroy, Garrett and Quill, Tyler J. and Siemons, Nicholas and Dijk, Gerwin and Marks, Adam and Cheng, Christina and Kleemann, Hans and Leo, Karl and Salleo, Alberto},
  journal = {Communications Materials},
  title = {Electron--ion coupling breaks energy symmetry in bistable organic electrochemical transistors},
  volume = {6},
  number = {1},
  pages = {241},
  year = {2025}
}

@article{wu2024bridging,
  title = {Bridging length scales in organic mixed ionic--electronic conductors through internal strain and mesoscale dynamics},
  author = {Wu, Ruiheng and Meli, Dilara and Strzalka, Joseph and Narayanan, Suresh and Zhang, Qingteng and Paulsen, Bryan D. and Rivnay, Jonathan and Takacs, Christopher J.},
  journal = {Nature Materials},
  volume = {23},
  number = {5},
  pages = {648--655},
  year = {2024},
  publisher = {Nature Publishing Group UK London}
}

@article{hohenberg1977theory,
  title = {Theory of dynamic critical phenomena},
  author = {Hohenberg, Pierre C. and Halperin, Bertrand I.},
  journal = {Reviews of Modern Physics},
  volume = {49},
  number = {3},
  pages = {435},
  year = {1977},
  publisher = {APS}
}

@article{cahn1958free,
  title = {Free energy of a nonuniform system. {I}. {I}nterfacial free energy},
  author = {Cahn, John W. and Hilliard, John E.},
  journal = {The Journal of Chemical Physics},
  volume = {28},
  number = {2},
  pages = {258--267},
  year = {1958},
  publisher = {American Institute of Physics}
}

@article{bongartz2024bistable,
  title = {Bistable organic electrochemical transistors: enthalpy vs. entropy},
  author = {Bongartz, Lukas M. and Kantelberg, Richard and Meier, Tommy and Hoffmann, Raik and Matthus, Christian and Weissbach, Anton and Cucchi, Matteo and Kleemann, Hans and Leo, Karl},
  journal = {Nature Communications},
  volume = {15},
  number = {1},
  pages = {6819},
  year = {2024},
  publisher = {Nature Publishing Group UK London}
}

@article{ghosh2020excitons,
  title = {Excitons and polarons in organic materials},
  author = {Ghosh, Raja and Spano, Frank C.},
  journal = {Accounts of Chemical Research},
  volume = {53},
  number = {10},
  pages = {2201--2211},
  year = {2020},
  publisher = {ACS Publications}
}

@article{metropolis1953equation,
  title = {Equation of state calculations by fast computing machines},
  author = {Metropolis, Nicholas and Rosenbluth, Arianna W. and Rosenbluth, Marshall N. and Teller, Augusta H. and Teller, Edward},
  journal = {The Journal of Chemical Physics},
  volume = {21},
  number = {6},
  pages = {1087--1092},
  year = {1953},
  publisher = {American Institute of Physics}
}

@misc{bongartz2025rustlatticesimulator,
  author = {Bongartz, Lukas M.},
  title = {An interactive lattice gas simulation tool},
  howpublished = {\url{https://lukasbongartz.github.io/rust-lattice-simulator/web-deploy/}},
  year = {2025}
}

@article{onsager1944crystal,
  title = {Crystal Statistics. I. A Two-Dimensional Model with an Order-Disorder Transition},
  author = {Onsager, Lars},
  journal = {Physical Review},
  volume = {65},
  number = {3--4},
  pages = {117--149},
  year = {1944}
}

@article{ferrenberg1991critical,
  title = {Critical behavior of the three-dimensional {Ising} model: A high-resolution {Monte Carlo} study},
  author = {Ferrenberg, Alan M. and Landau, D. P.},
  journal = {Physical Review B},
  volume = {44},
  number = {10},
  pages = {5081--5091},
  year = {1991}
}

@article{bisquert2023hysteresis,
  title={Hysteresis in organic electrochemical transistors: distinction of capacitive and inductive effects},
  author={Bisquert, Juan},
  journal={The Journal of Physical Chemistry Letters},
  volume={14},
  number={49},
  pages={10951--10958},
  year={2023}
}

@article{ji2021mimicking,
  title = {Mimicking associative learning using an ion-trapping non-volatile synaptic organic electrochemical transistor},
  author = {Ji, Xudong and Paulsen, Bryan D. and Chik, Gary K. K. and Wu, Ruiheng and Yin, Yuyang and Chan, Paddy K. L. and Rivnay, Jonathan},
  journal = {Nature Communications},
  volume = {12},
  number = {1},
  pages = {2480},
  year = {2021},
  publisher = {Nature Publishing Group UK London}
}

@book{chandler1987introduction,
  author    = {D. Chandler},
  title     = {Introduction to Modern Statistical Mechanics},
  publisher = {Oxford University Press},
  address   = {New York},
  year      = {1987},
  pages     = {67}
}

@article{newman2000efficient,
  title={Efficient Monte Carlo Algorithm and High-Precision Results for Percolation},
  author={Newman, MEJ and Ziff, RM},
  journal={Physical Review Letters},
  volume={85},
  number={19},
  pages={4104--4107},
  year={2000}
}

@article{wang2013bond,
  title={Bond and site percolation in three dimensions},
  author={Wang, Junfeng and Zhou, Zongzheng and Zhang, Wei and Garoni, Timothy M and Deng, Youjin},
  journal={Physical Review E},
  volume={87},
  number={5},
  pages={052107},
  year={2013}
}

\end{document}